\documentclass[twocolumn,notitlepage,superscriptaddress, nofootinbib,nobibnotes, aps,prd,10pt]{revtex4-1}% secnumarabic, 
\usepackage{amsmath,amssymb, fp}  
\usepackage{xcolor}  
\usepackage{soul} 
\usepackage{booktabs}
\definecolor{linkcolor}{RGB}{7,94,84}  %teal dark green
\usepackage[	colorlinks=true,    	
			pdfstartview=FitV,
			linkcolor= linkcolor,
			citecolor= linkcolor,
			urlcolor= linkcolor,
			linktoc=page,
			hyperindex=true,
			hyperfigures=true,  
			breaklinks=true]
			{hyperref}
\usepackage{dblfloatfix}    
\usepackage{balance} 
\usepackage{lastpage}  
\usepackage{mathtools}
\usepackage{epsfig,rotating,pifont}
\usepackage{graphicx}
\graphicspath{{./figures/}}	
\usepackage[caption=false]{subfig}
\usepackage{placeins}
\usepackage{ragged2e} 
\usepackage{etoolbox}
\usepackage{changepage}   % for the adjustwidth environment
\usepackage{pgfplots}
\usetikzlibrary{pgfplots.groupplots}
\pgfplotsset{compat=1.3}
\usepackage{tikz}  
\usetikzlibrary{arrows,shapes,positioning}
\usetikzlibrary{decorations.markings,decorations.pathmorphing,decorations.pathreplacing,decorations.text}
\usetikzlibrary{arrows,calc,patterns,shapes.geometric}
\usepackage{fancybox}
\usepackage{verbatim}
\usepackage{multirow}
\usepackage{titlesec}

\newsavebox\CBox

\def\beq{\begin{eqnarray}}
\def\eeq{\end{eqnarray}}

\def\la{\langle }
\def\ra{\rangle }

\def\lb{\label}

\def\O{\mathcal{O}}

\newcommand{\be}{\begin{equation}}
\newcommand{\ee}{\end{equation}}
\newcommand{\bea}{\begin{eqnarray}}
\newcommand{\eea}{\end{eqnarray}}
\newcommand{\bg}{\begin{gather}}

\newcommand{\bseq}{\begin{subequations}}
\newcommand{\eseq}{\end{subequations}}

\def\tr{{\rm Tr}}

\def\be{\begin{eqnarray}}
\def\ee{\end{eqnarray}}
\def\lb{\label}

\def\nn{\nonumber}
\def\hs{\hspace{1pt}}

\definecolor{Green}{RGB}{147,162,153}
\definecolor{Green2}{RGB}{26,148,49}
\definecolor{BrownL}{RGB}{173,143,103}
\definecolor{Red}{RGB}{210,83,60}
\definecolor{BrownD}{RGB}{114,96,86}
\definecolor{GreyD}{RGB}{76,90,106}
\definecolor{GreyB}{RGB}{128,141,160}
\definecolor{Maroon}{RGB}{121,70,61}
\definecolor{Blue}{RGB}{148,184,210}
\definecolor{Blue2}{RGB}{108,144,170}
\definecolor{Blue3}{RGB}{42, 107, 172}
\definecolor{BB}{RGB}{128,184,220}  
\newcommand{\red}[1]{#1}

\newsavebox\foobox
\begin{document}

\title{Entanglement of skeletal regions}  

\author{Cl\'ement Berthiere}
%\email{clement.berthiere@umontreal.ca}
\affiliation{Universit\'{e} de Montr\'{e}al, C.\,P.\,6128, Succursale Centre-ville, Montr\'eal, QC, Canada, H3C 3J7}
\affiliation{Centre de Recherches Math\'{e}matiques, Universit\'{e} de Montr\'{e}al, Montr\'{e}al, QC, Canada, H3C 3J7}

\author{William Witczak-Krempa}
%\email{w.witczak-krempa@umontreal.ca}
\affiliation{Universit\'{e} de Montr\'{e}al, C.\,P.\,6128, Succursale Centre-ville, Montr\'eal, QC, Canada, H3C 3J7}
\affiliation{Centre de Recherches Math\'{e}matiques, Universit\'{e} de Montr\'{e}al, Montr\'{e}al, QC, Canada, H3C 3J7}

\date{\today} 
 
\begin{abstract}\vspace{3pt}
%\begin{center}\textbf{\abstractname}\end{center}\vspace{-3pt} 
%
The entanglement entropy (EE) encodes key properties of quantum many-body systems. It is usually calculated for subregions of finite volume (or area in 2d). 
In this work, we study the EE of skeletal regions that have \textit{no} volume, such as a line in 2d. We show that skeletal entanglement displays new behavior compared to its bulk counterpart, and leads to distinct  universal  quantities. We provide non-perturbative bounds for the skeletal area-law coefficient of a large family of quantum states.
We then explore skeletal scaling for the toric code, conformal bosons and Dirac fermions,
Lifshitz critical points, and Fermi liquids. We discover signatures including skeletal topological EE, novel \red{corner} terms, and strict area-law scaling for metals. \red{These findings suggest that skeletal entropy serves as a measure for the range of entanglement.}
We discuss the possibility of a continuum description involving the fusion of defect operators. 
Finally, we outline open questions relating to other systems, and measures such as the logarithmic negativity. 

\vspace*{-2pt}
$\,$

\end{abstract}

\maketitle  

\makeatletter
\makeatother
%\tableofcontents

Recent years have witnessed significant endeavors to further our understanding of the entanglement properties of quantum many-body systems \cite{Amico:2008aa,Calabrese_2009}. In quantifying entanglement, several measures have emerged~\cite{Plenio:2007zz,RevModPhys.81.865}, each with its scope of applicability, and among which the EE represents an important instance \cite{Calabrese:2004eu,Calabrese:2009qy,Casini:2009sr,RevModPhys.82.277}. The EE follows from partitioning a system into two complementary subsystems, say $A$ and $B$, and evaluating the von Neumann entropy $S=-\tr\hs\rho_A\log\rho_A$ of the reduced density matrix $\rho_A$ of one subsystem. The R\'enyi generalization is defined as $S_n=\frac{1}{1-n}\log\tr\hs\rho_A^n$, recovering the EE in the limit $n\rightarrow1$. 
The characteristic size of the subsystems is usually taken far larger than the microscopic length scale, which can be a lattice spacing for lattice models or a UV cutoff in a quantum field theory. For extended systems in $d$ spatial dimensions, one is thus generally interested in the EE of subsystems with volume. In this Letter, we initiate the systematic study of entanglement properties of subsystems possessing \textit{no} volume, i.e.~having spatial codimension $p$ with $1\le p\le d$\red{, as illustrated in Fig.~\ref{figZ} in 2d for $p=1$}. We dub such subsystems ``skeletal''. 
For 1d systems, single-site entanglement has been shown to provide a diagnostic of quantum phase transitions \cite{PhysRevLett.93.086402,PhysRevLett.95.056402,PhysRevLett.95.196406}, and has found successful use in the context of quantum impurity models \cite{2007JSMTE..08....3S,2009JPhA...42X4009A}. In higher dimensions, although studies have been performed on certain specific models/subregions \cite{Hamma:2004vdz,2012PhRvB..86u4203K,2014PhRvB..89q4202K,2015PhRvB..91o5145L,2015PhRvB..92k5126L}, much less is known about 
entanglement measures for skeletal regions.

We investigate groundstates of quantum systems in $d>1$ through the lens of EE of skeletal regions, unveiling new subleading scaling behaviors of the EE, associated with new universal quantities. 
%On general grounds, one expects the (R\'enyi) EE of codimension-$p$ regions to scale as $S\sim\beta \hs2\ell^{d-p}$.
We mainly work with 2d systems and codimension-1 skeletal regions. For all cases considered, the EE takes the general form
\be\lb{EE}
S = \beta\hs 2\ell -\gamma - s\,,
\ee
where $\ell$ is the length of the skeletal region (or half its perimeter as defined on the lattice) taken to be far larger than the short-distance cutoff (set to 1 without loss of generality). The first term is the area-law, with coefficient $\beta$. % thus being nonuniversal (it depends on the regularization scheme). 
The first subleading correction parametrized by $\gamma$ contains more useful information about the quantum state; it can be universal, and finite or divergent with $\ell$ depending on the geometry of the skeletal region. Finally, $s$ denotes subleading terms that vanish in the large-$\ell$ limit. We will see that the scaling of $s$ with respect to $\ell$ strongly depends on the theory, and that $s$ can capture universal quantities as well.

\begin{figure}[t]
\centering\vspace{4pt}
\includegraphics[scale=0.97]{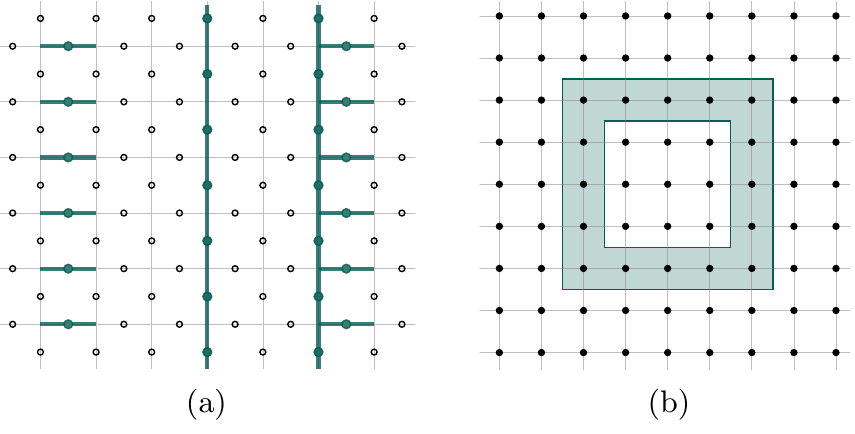}\vspace{-4pt}
\caption{Typical skeletal regions on 2d square lattices for\linebreak (a) the toric code \red{(ladder, chain, and chain-ladder)}, and \linebreak (b) Dirac and scalar discretized CFTs.}
\lb{figZ}
\vspace{-4pt}
\end{figure}

A general trend we observe for skeletal regions is that their EE is generally `weaker' than that of the corresponding subregions with volume, \red{which can be understood from the significant reduction of degrees of freedom in the subregion per area of interface compared to a bulk region, thus limiting the amount of entanglement.}
We illustrate this by putting \red{non-perturbative} bounds on the area-law coefficient for general \red{many-body} lattice systems. \red{It also translates into weaker subleading corrections, as we exemplify with topological entanglement entropy in the toric code, or cusps in certain CFTs.
Interestingly, we find that the strength of skeletal corrections is an indicator of the spatial range of entanglement, with systems known to possess shorter-range entanglement giving rise to maximally strong skeletal corrections.}

\pagebreak[5]

%\medskip\smallskip
\noindent\textit{\textbf{Bounding the skeletal area-law.}}
Consider a quantum state that satisfies the entanglement area-law. For concreteness, we work on a $d$-dimensional hypercubic lattice of $\ell^d$ sites. 
Let three subsystems $A,B,C$ span the whole lattice in every direction except one, $y$, along which the subsystems are $w_{A,B,C}\ge1$ sites wide and such that $A$ is adjacent to $B$ and $C$, \red{as depicted for 2d in the inset of Fig.\,\ref{figY}}.
For such subsystems, one then expects the EE to scale as $S = \beta_w 2\ell^{d-1} + \cdots$ at large $\ell$, where $\beta_w$ is the area-law coefficient for a subsystem of width $w$. \red{A \mbox{codimension-1} minimal skeletal region corresponds to $w\!=\!1$; we denote the corresponding skeletal coefficient $\beta\equiv\beta_1$.} 
Assuming a translation invariant state along $y$, and using the positivity of mutual information $S_A+S_B-S_{A\cup B}\ge0$ \red{for $w_B=1$ and $w_A=1,2,\cdots$}, we obtain a lower bound on the \red{minimal} skeletal coefficient: $\beta\ge \beta_w/w$. 
%\red{This bound follows from successive applications of positivity of mutual information for two adjacent skeletal regions $A$ and $B$ as described above, setting $w_{A}=1$ and for $w_B\ge1$.}
We can put an upper bound on $\beta$ using the strong subadditivity inequality \red{$S_{A\cup B\cup C}+S_A\le S_{A\cup B}+S_{A\cup C}$ for $w_B=w_C=1$ and $w_A=1,2,\cdots$,} finding $\beta \leq w \beta_w - (w-1)\beta_{w+1} \,(\leq \beta_w)$. %\red{The first inequality is obtained by setting $w_B=w_C=1$ and applying SSA successively for $w_A\ge1$.} 
We thus have the following bounds on the skeletal area-law prefactor $\beta$:
\be\label{areabounds}	
\beta_w/w  \leq \beta \leq w \beta_w - (w-1)\beta_{w+1} \leq \beta_w\,.\;\;
\ee
The tightest bounds are for $w=2$. Note that the upper bound in \eqref{areabounds} implies that the skeletal coefficient $\beta$ is always less or equal to the asymptotic coefficient $\beta_\infty$ for subsystems with volume in the thermodynamic limit. Such bounds can be straightforwardly 
generalized to cases where translation symmetry is reduced, and
to codimension-$p$ regions.

\medskip\smallskip
\noindent\textit{\textbf{Toric code.}} Let us now consider the entanglement properties of skeletal regions in the groundstate of a gapped $\mathbb{Z}_2$ quantum spin liquid: the 2d toric code \cite{Kitaev:1997wr}.  
The Hamiltonian of this strongly interacting spin-$1/2$ model is $-\sum_s A_s -\sum_p B_p$, where 
$A_s=\prod_{j\in s}\sigma_j^x$ is a star operator acting on the four bonds touching a vertex on the square lattice, while $B_p=\prod_{j\in\partial p}\sigma_j^z$ involves a product around a unit plaquette; see Fig.\,\hyperref[figZ]{\ref{figZ}(a)}. We work with one of the four groundstates on the square torus, %$|\xi_{00}\rangle$, 
which is an equal weight superposition of all closed-string states. %(each being a product state of up/down spins where the up spins form closed loops). 
%
%For our skeletal regions, 
Let us first take lines of spins that go around the vertical direction, the simplest cases being the \red{ladder and chain,  
shown in order} in Fig.\,\hyperref[figZ]{\ref{figZ}(a)}. The EE for these two subregions~\cite{Hamma:2004vdz} obeys the general relation \eqref{EE} (here taking the logarithm base 2) with $\beta=1/2$, and $\gamma=1$ for the chain, while $\gamma=0$ for the ladder. For all regions, $s=0$.
It is natural that the area-law coefficients coincide, but somewhat surprising that the constant terms do not. 
For the chain, $\gamma$ yields the topological entanglement entropy (TEE) $\log_2\mathcal D$, with the total quantum dimension of the toric code $\mathcal D=\sqrt 4$. This is the same value of $\gamma$ as for a non-skeletal strip~\cite{Zhang2012}. 
If we join the chain and ladder to form another skeletal region, Fig.\,\hyperref[figZ]{\ref{figZ}(a)}, 
we find a sudden increase of the area-law coefficient to its thermodynamic value $\beta=1$, while $\gamma=1$ (Appendix~\ref{SM-toric}).
As we further make the region wider,  we find that the EE remains constant.
We thus have a sharp transition from the skeletal to the regular regime, in contrast to the slow convergence that we will discover in quantum critical states below. Further, the chain and ladder regions saturate the lower bound in \eqref{areabounds}, adapted to the geometry in Fig.\,\hyperref[figZ]{\ref{figZ}(a)}.   

In order to put the skeletal TEE on firmer footing, we consider the Levin-Wen \cite{2006PhRvL..96k0405L} subtraction scheme $S(\square)-S(\sqcup) -S(\sqcap)+S(\vcenter{\hbox{\scalebox{1.05}{\hspace{-2.3pt}\rotatebox{90}{$-$}\hspace{-1.8pt}\rotatebox{90}{$-$}}\hspace{-2.5pt}}})$. When the four sides are either a chain, ladder or chain-ladder,
we find that the linear combination gives $-1$. Starting with the chain-ladder-chain, we obtain the thermodynamic result, $-2$. It would be interesting to investigate the skeletal-to-thermodynamic crossover with non-fixed-point wavefunctions, and other topological states.

\begin{figure}[t]
\centering\vspace{4pt}
\includegraphics[scale=0.999]{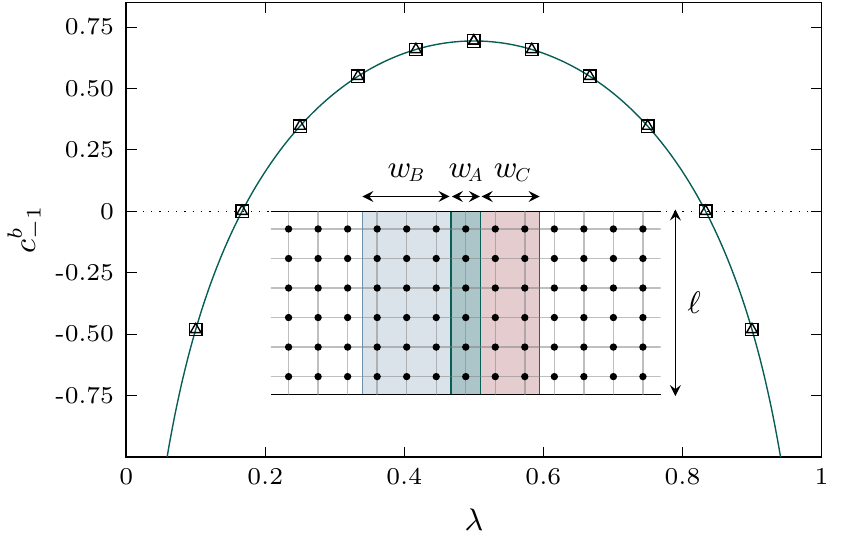}\hspace{10pt}%\vspace{-5pt}
\caption{Universal coefficient $c_{-1}^b$ (see \eqref{sub}) as a function of the twist parameter $\lambda$ for the complex boson. Squares show numerical data points for R\'enyi index $n=1$, and triangles for $n=2$. Numerical results are compared to $c_{-1}^b = \log(2\sin\pi\lambda)$ shown in green. \textbf{Inset}: \red{skeletal regions $A,\,B,\,C$ of widths $w_{A,B,C}$ and spanning the whole lattice in the vertical direction. Example of configurations used to derive the bounds \eqref{areabounds} on the area-law coefficient. The numerical results for $c_{-1}^b$ were obtained for the region $A$ consisting in a circle of $\ell$ sites on the infinitely long cylinder, where top and bottom are identified.}}
\lb{figY}
%\vspace{-3pt}
\end{figure}

\medskip\smallskip
\noindent\textit{\textbf{Dirac \& scalar CFTs.}}
We next consider two quantum critical theories with emergent Lorentz and scale invariance (CFTs), namely the free complex boson %with Lagrangian density $\frac{1}{2}\partial_\mu\phi^*\partial^\mu\phi$, 
and the free Dirac fermion. %, $i\bar \psi \gamma^\mu\partial_\mu\psi$.  
We compute the corresponding groundstate EEs for skeletal regions of different shapes, unveiling new universal contributions.  We discretize both theories on the 2d square lattice (see Appendix \ref{SM1}), and calculate the spectrum of the reduced density matrix from the two-point functions restricted to the subregion~\cite{Casini:2009sr}. We first obtain the skeletal area-law coefficients $\beta_w$ for strips of widths $w=1,2,3$, calculated semi-analytically to arbitrary precision. They obey the bounds \eqref{areabounds}, see Appendix \ref{SM2}. 

We now analyze the subleading terms, starting with the simplest case % where semi-analytical results can be obtained: 
of a circle of $\ell$ lattice sites on the infinitely long cylinder. We impose twisted boundary conditions in the compact direction, $\mbox{$\phi(x,y+\ell)$}=e^{-2\pi i \lambda}\phi(x,y)$ for bosons, and similarly for Dirac fermions, where $\lambda\in(0,1)$. 
For both the complex boson and Dirac fermion we find that the (R\'enyi) EE has the form \eqref{EE}, see Appendix \ref{SM2}. %The area-law coefficients can be calculated semi-analytically on the lattice to arbitrary precision. 
The universal constant $\gamma$ is found to vanish, independently of the Rényi index $n$, for both theories. %the free boson and free Dirac fermion. 
In contrast, the form of the subleading term $s$ in \eqref{EE} strongly depends on the theory. We find (see Appendix \ref{SM2})
\be\lb{sub}
s =
\begin{cases}
\displaystyle\;\frac{c^b_{-1}}{\log \ell} - \frac{c_{-2}^b}{(\log\ell)^2}+\cdots\,, &\;\; {\rm boson}\,,\vspace{5pt}\\
\displaystyle\; c_{-1}^f\frac{\log\ell}{\ell^2} -\frac{c_{-2}^f}{\ell^2} + \cdots\,, &\;\; {\rm fermion}\,.\quad
\end{cases}
\ee
The difference in behavior of $s$ at large $\ell$ between bosons and fermions is striking. For free bosons, $s$ vanishes very slowly, as $1/\log\ell$,
while for free Dirac fermions $s$ falls off much faster as $(\log \ell)/\ell^2$. 
\red{We interpret this difference between boson and fermion CFTs as an indicator that fermions possess longer-range entanglement compared with bosons, so that their universal contributions cannot be readily probed with skeletal regions. Below, we shall present a system of nonrelativistic bosons with maximally strong skeletal universal terms, hence shortest-range entanglement.}
The (inverse) logarithmic scaling of $s$ for bosons protects the coefficient $c_{-1}^b$ from being spoiled by the short-distance cutoff, it is thus universal. In contrast, no such quantity appears in $s$ for Dirac fermions. %The corresponding universal coefficient $c_{-1}^b$ for the free complex boson can be extracted numerically. 
We numerically extract $c_{-1}^b$ for the free complex boson. It depends on the twist parameter $\lambda$, and perfectly fits the relation $c_{-1}^b = \log(2\sin\pi\lambda)$,
as can be seen in Fig.\,\ref{figY}. Moreover, we observe that $c_{-1}^b$ is independent of the Rényi index $n$. 
Quite surprisingly, $c_{-1}^b$ matches the universal constant term in the Rényi EE of the bipartite infinite cylinder \cite{Chen:2016kjp}, up to an overall factor that cancels the $n$ dependence. 

Let us now concentrate on the complex boson, as it presents richer universal features than the Dirac fermion. An important class of skeletal regions consists in broken lines presenting cusps. The intersection of two straight segments in the bulk forms a cusp with opening angle $\theta$, see Fig.\,\hyperref[figZ]{\ref{figZ}(b)}. When the system has a boundary, a straight segment can intersect that boundary and form another kind of cusp. We call the former kind ``bulk cusp" and the latter ``boundary cusp''. \red{The (Rényi) EE of skeletal regions with boundary cusps is given by \eqref{EE}, with  subleading corrections} % due to the presence of cusps
\be\lb{cusps}
\gamma = \sum_i b_{\rm sk}(\theta_i) \log\log\ell +c_0\,,\;
%\gamma_{\rm cusps} =
%\begin{cases}
%\displaystyle\, \sum_i a_{\rm sk}(\theta_i) \,, &{\rm bulk\,\,cusps}\,,\vspace{4pt}\\
%\displaystyle\,\sum_i b_{\rm sk}(\theta_i) \log\log\ell +c_0\,, &{\rm boundary\,\,cusps}.\quad
%\end{cases}
\ee
\red{while the contributions of bulk cusps is $\O(1)$, $\gamma=\sum_i a_{\rm sk}(\theta_i)$ (we refer the reader to Appendix \ref{SM3} for details).
%The EE of skeletal regions with cusps thus displays novel scaling behavior.
Skeletal entanglement thus displays novel bulk and boundary cusps terms.}
Indeed, as it is well-known \cite{Casini:2006hu,Casini:2008as,pitch,Bueno:2015rda,Faulkner:2015csl,Berthiere:2016ott,FarajiAstaneh:2017hqv,Seminara:2017hhh,Berthiere:2018ouo,Berthiere:2019lks,Rozon:2019evk},
2d regions with bulk and/or boundary corners show a 
\pagebreak[3]
logarithmic divergence at large $\ell$ in gapless critical systems. 
We observe that the coefficient $c_{-1}^b$ (see \eqref{sub}) is still universal for bulk cusps, but it is not for boundary cusps. On the lattice, $\mathcal{O}(1)$ terms in the EE may be polluted by the leading area-law. Henceforth we focus on the unambiguous universal quantities that are $c_{-1}^b$ for bulk cusps and $b_{\rm sk}(\theta)$ for boundary cusps.

We have computed the EE on the infinite 2d lattice for several polygonal skeletal regions. %, such as squares, isosceles rectangle triangles and octagons. 
Numerical results for $c_{-1}^b$ are reported in \mbox{Table\hspace{3pt}\ref{tab}} of Appendix \ref{SM3}. They suggest that the contributions of the bulk cusps are additive, i.e. $-c_{-1}^b=\sum_i \tilde{a}_{\rm sk}(\theta_i)$ where $\tilde{a}_{\rm sk}(\theta)$ is a new bulk cusp function. % distinct from $a_{\rm sk}(\theta)$ in \eqref{cusps}. 
We observed that $c_{-1}^b$ does not depend on the R\'enyi index $n$.
%Note that the bulk cusp functions we discuss here are different than the well-known corner function of 2d regions mentioned earlier.

\red{For boundary cusps, the coefficient of the subleading double-logarithmic scaling term in the EE is universal, see \eqref{cusps}. We have calculated $b_{\rm sk}(\theta)$ for Dirichlet (D) and Neumann (N) conformal boundary conditions, and angles $\theta=\pi/2,\,\pi/4$; see Appendices \ref{SM2} and \ref{SM3}. We remark that $b_{\rm sk}(\theta)$ is independent of $n$. More importantly, we observe the relation $b_{\rm sk}^{(N)}(\theta)=-b_{\rm sk}^{(D)}(\theta)$, which we conjecture to hold for any angle $\theta\in(0,\pi)$, see \cite{Berthiere:2019lks}. Neumann and Dirichlet contributions differ only by a sign, as does the boundary central charge  $\mathfrak{a}$ in $2+1$ dimensions~\cite{Fursaev:2016inw,Berthiere:2019lks} for free scalars.
We thus conjecture that for free fields, the boundary cusp function satisfies $b_{\rm sk}(\theta) = \mathfrak{a} f(\theta)$, with $f(\theta)$ being a function of the opening angle only. This would imply that it vanishes identically for Dirac fermions with mixed boundary conditions since $\mathfrak{a}=0$ in that case.}
%The double logarithm appearing in \eqref{cusps} for boundary cusps is particularly interesting. 
%We computed its universal coefficient for $\theta=\pi/2$ (analytically) and $\pi/4$ (numerically) for the free complex boson with Dirichlet (D) and Neumann (N) boundary conditions, see Appendices \ref{SM2} and \ref{SM3}. 
%We note that $b_{\rm sk}(\theta)$ is independent of $n$. More importantly we remark that $b_{\rm sk}^{(N)}(\theta)=-b_{\rm sk}^{(D)}(\theta)$ for both angles considered. We believe this relation to hold for any angle $\theta\in(0,\pi)$, and we actually do expect it from the bulk-boundary relation of \cite{Berthiere:2019lks} which relates the R\'enyi entropies of certain theories with and without a boundary. %\Blue{[CB: should we delete this paragraph?] Applied to skeletal regions, the bulk-boundary relation indeed implies $b_{\rm sk}^{(D)}(\theta)+b_{\rm sk}^{(N)}(\theta)=0$ since there are no double logarithmic divergence for bulk cusps.} 
%Further, we conjecture that for free fields, the boundary cusp function $b_{\rm sk}(\theta) = \mathfrak{a} f(\theta)$ is proportional to the boundary central charge $\mathfrak{a}$ in $2+1$ dimensions~\cite{Fursaev:2016inw,Berthiere:2019lks}, with $f(\theta)$ being a function of the opening angle only. This would imply that $b_{\rm sk}(\theta)$ vanishes identically for Dirac fermions with mixed boundary conditions since $\mathfrak{a}=0$ in that case ($\mathfrak{a}=\pm1$ for D/N free real scalars). %It would be interesting to test this conjecture. %, e.g., against further numerical calculations on the lattice.

Finally, to probe the universality of the subleading terms in the EE for the free complex boson, we considered two %one-parameter 
families of \red{next-to-nearest neighbor} deformations of the lattice boson theory%. The first one is a next-to-nearest neighbor deformation, while the second interpolates between the relativistic boson (with dynamical exponent $z=1$) and the Lifshitz ($z=2$) boson
, see Appendix \ref{SM4}.
%We compute the EE of a twisted circle on the infinitely long cylinder, and of a line intersecting the Dirichlet boundary of an infinite strip at angle $\pi/2$. 
We find that, in the scaling limit, the subleading properties of the EE are robust against both deformations--an indication of the universal nature of the information encoded in $\gamma$ and $s$.% (see \eqref{EE} and \eqref{sub}).

\medskip\smallskip%\vspace{2pt}

\noindent\textit{\textbf{Lifshitz criticality.}}
Lifshitz field theories represent a class of nonrelativistic QFTs which exhibit anisotropic scaling between space and time, with dynamical exponent $z\neq1$. 
In $d+1$ dimensions, free \mbox{Lifshitz scalar} theories with positive even $z$ enjoy many interesting features. The Hamiltonian for the real massless noncompact $z = 2$ Lifshitz \mbox{quantum critical} boson reads $\frac{1}2\int d^dx(\pi^2 + (\nabla^2\phi)^2)$. %, where $\kappa$ is a dimensionless parameter. 
The corresponding groundstate wavefunctional--of \mbox{Rokhsar-Kivelson-type--is} given in terms of the partition function of a \red{\emph{local}} \mbox{$d$-dimensional} free Euclidean scalar field, a CFT \cite{Ardonne:2003wa,Angel-Ramelli:2019nji}.  
In 2d, the celebrated Fradkin-Moore formula \cite{Fradkin:2006mb} gives the bipartite (R\'enyi) EE for the Lifshitz groundstate, 
\mbox{$S=-\log(Z^{AB}_D/Z_F)$,}
up to nonuniversal terms, where $Z^{AB}_D$ is the $\text{CFT}_2$ partition function of configurations with Dirichlet boundary conditions on the boundary of $A$ ($\partial A=\partial B$), and $Z_F$ is the partition function of free configurations. %In flat space, 
This formula generalizes to all dimensions and positive even $z$~\cite{Angel-Ramelli:2019nji,Boudreault:2021pgj}, and \red{we show that it} can be applied to skeletal regions.

As an illustration, let $A$ be a circle on the infinite cylinder. Then $B=B_1\cup B_2$ is the union of two semi-infinite cylinders $B_1$ and $B_2$ separated by $A$. The EE of $A$ is thus equivalent to the EE of the \mbox{bipartition} $\{B_1,B_2\}$, which has been calculated in \cite{Chen:2016kjp} for the \mbox{complex} \mbox{Lifshitz} boson with twisted boundary condition. The EE displays a universal constant correction, which reads \mbox{$\gamma = \log(2\sin\pi\lambda)$}, \red{in contrast with relativistic bosons and fermions where $\gamma=0$}.
We have  numerically verified this on the lattice. %We have also checked that the relation for skeletal area-law coefficients $\beta_w\simeq \beta_\infty-\kappa/(2w)$ holds to $0.56\%$ accuracy down to $w=1$.

More generally, for Gaussian groundstates of $(d+1)$-dimensional free noncompact Lifshitz theories with positive even dynamical exponent $z$, the (R\'enyi) EE of a spatial codimension-one skeletal region that also defines a bipartition of the system is equivalent (i.e.~up to nonuniversal terms) to the EE for that bipartition. %In other words, if a skeletal region can be seen as the entangling surface that separates two complementary regions, the EE of the entangling surface itself (i.e. the skeletal region) is the EE across the entangling surface.
\red{This translates into subleading terms being as strong as for bulk regions, in sharp contrast with what we found for CFTs.}
The reason behind this is the local nature of the Lifshitz groundstate, which translates into entanglement ``localization''.

Let us finally consider a skeletal region that does not define a bipartition of the system. Take $A$ to be an open segment of length $\ell$. While $A$ has no area, its complementary domain $B$ does, and presents two corners of opening angle $\theta=2\pi$. Using the classic work of Cardy and Peschel \cite{CARDY1988377} that gives the CFT$_2$ free energy on domains with boundaries, these corners produce a logarithmic term in the EE \eqref{EE}, i.e. $\gamma = 2\times(-1/8)\log\ell$. %, each end-point of the segment thus contributing $-1/8$ to the coefficient. 
Our field theoretical prediction is in perfect agreement with the lattice numerics, further showing that $s=\O(\ell^{-1})$. We also find, numerically, the same $\gamma$ for an open segment in $d=3$ dimensions (cubic lattice).

The (disorder-averaged) EE of certain skeletal regions for the bond-diluted quantum Ising model displays universal logarithmic corrections \cite{2012PhRvB..86u4203K,2014PhRvB..89q4202K}. These can be obtained applying Cardy-Peschel formula, thus sharing similar structures with the EE in $z=2$ Lifshitz theory.

\medskip\smallskip\smallskip
\noindent\textit{\textbf{Metals.}}
Systems with a Fermi surface are known to violate the area-law of EE in the form of a multiplicative logarithmic correction \cite{Wolf:2006zzb,Gioev:2006zz,2006PhRvA..74b2329B,2006PhRvB..74g3103L,Swingle:2009bf}: $S\sim\ell^{d-1}\log\ell$, for a region $A$ with volume of characteristic size $\ell$. Now, for a skeletal region, how would the EE scale for such systems?
To answer this question, take a simple geometry where the region $A$ is a strip of width $w$ in $d$ spatial dimensions with translation invariance in $d-1$ transverse dimensions. We compactify these transverse directions to have size $\ell$. The EE then scales as $S\sim \ell^{d-1}\log w$ \cite{Wolf:2006zzb,Swingle:2009bf,Swingle:2010yi}. A skeletal region of fixed width $w\ll\ell$ will thus have an EE satisfying the area-law.

Let us now consider free spinless fermions hopping on a 2d square lattice between nearest neighbor sites, and with chemical potential $\mu$. %The groundstate is a Fermi sea where all states with energy less than $\mu$ are filled. 
Half-filling corresponds to $\mu=0$ where the Fermi surface is a square.
We take the 2d lattice to be infinite in one direction, while we impose twisted boundary conditions (with twist parameter $\lambda$) in the other direction. The skeletal region is a chain of $\ell$ lattice sites along the latter direction. %Given the symmetry, we can dimensionally reduce our problem of computing the EE of a circle of $\ell$ sites to that of the calculation of a single site entropy on $\ell$ decoupled 1d infinite lattice.
The square Fermi surface case ($\mu=0$) %, corresponding to a square Fermi surface,
is analytically tractable. The EE of a (twisted) circle is given by \eqref{EE} with $\beta=1/4$ and $\gamma=0$. 
For the subleading term we find $s=c_{-1}(\lambda)\,\ell^{-1}\log\ell + \O(\ell^{-1})$.
%, and we observe a parity effect with $\ell$, 
%\be
%c_{-1}= 2/3-4\lambda(1-\lambda)+(1-(-1)^\ell)(\lambda-1/4)  \,,\quad
%\ee
%for values of the twist parameter $\lambda\in[0,1/2]$. The range $\lambda\in[1/2,1]$ is obtained by changing $\lambda\rightarrow 1-\lambda$. 
%The coefficient $c_{-1}$ for $\ell$ even is invariant under that change, while that for $\ell$ odd is not. Interestingly, the odd coefficient is thus not a smooth function of $\lambda$.
%Note that Dirichlet boundary conditions are straightforwardly implemented, finding \eqref{EE} for the entropy with $\beta=1/4$, $\gamma=-1/2$ and $s = \log \ell/(6\ell) + \O(\ell^{-1})$. Parity effects can be observed as well for Dirichlet boundary conditions for general $\mu$, though not for $\mu=0$. \red{Interestingly, $\mu=1$ appears to be a special value. Indeed, the subleading term $s$ does not oscillate with $\ell$ even or odd. REMOVE?}

\begin{figure}[t]
\centering
\includegraphics[scale=1.02]{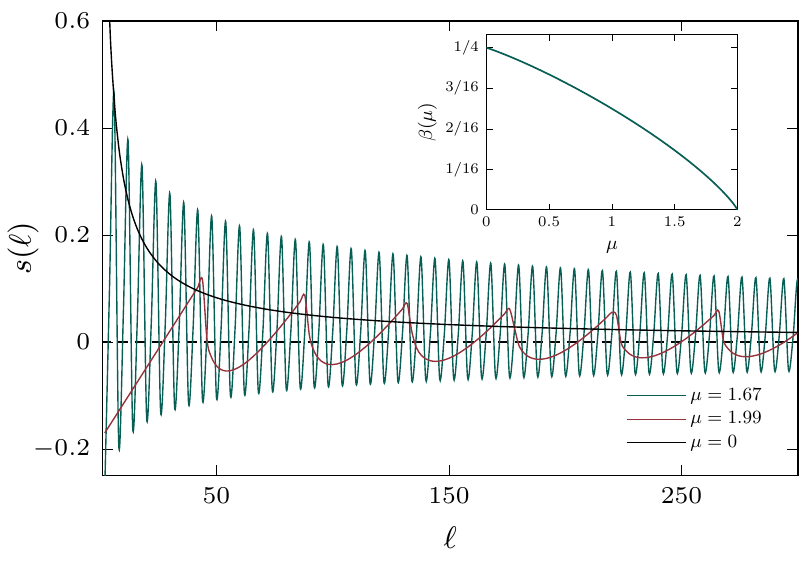}\vspace{-5pt}
\caption{Subleading term $s$ as a function of $\ell$ even for periodic boundary condition $\lambda=0$, and different values of the chemical potential $\mu$. \textbf{Inset}: Area-law coefficient $\beta$ as a function of $\mu$.}
\lb{figX}
\vspace{-3pt}
\end{figure}

For the more general case of a smooth Fermi surface, the area-law coefficient $\beta$ can be computed semi-analytically as a function of $\mu$, see Fig.\,\hyperref[figX]{\ref{figX}} (inset), and the constant $\gamma$ in the EE is found to vanish. \mbox{Figure \hyperref[figX]{\ref{figX}}} shows the subleading oscillating corrections encoded in $s$, \red{which decay with $\ell$ in contrast to bulk regions that have a linearly divergent correction~\cite{Murciano:2020lqq}}. 
Such oscillations in the EE of subsystems with volume are observed for the tight-binding model in 1d \cite{2010JSMTE..08..029C} and 2d \cite{Murciano:2020lqq}. However, their behaviors are quantitatively different than what we observe for skeletal regions. When $\mu$ tends to 2, i.e. the Fermi surface becomes circular as it shrinks to a point, parity effects disappear and the period of oscillation increases. In that regime, the period of oscillation approaches $2\pi/k_F$, where $k_F$ is the Fermi momentum.
For $\mu>0$, the decrease of the amplitude \mbox{is slower than for $\mu=0$, as can be seen in Fig.\,\hyperref[figX]{\ref{figX}}.}

\medskip\smallskip
\noindent\textit{\textbf{Discussion.}}
We have initiated the systematic study of entanglement of skeletal regions, i.e. subregions with no volume, through the lens of entanglement entropy. We showed that skeletal entanglement displays new behavior compared to standard bulk subregions. \red{This leads to distinct universal quantities, which, interestingly, are independent of the R\'enyi index, in stark contrast to what is found for non-skeletal regions.}
%It is interesting to note that these universal subleading terms are independent of the R\'enyi index, in stark contrast to what is found for non-skeletal regions. 
\red{We have also argued that skeletal entanglement serves as an indicator of the spatial range of entanglement, through the strength of
subleading skeletal corrections.}

A future line of inquiry concerns the definition of EE of skeletal regions in continuum field theories such as CFTs. Though we have presented results for the continuum $z=2$ Lifshitz theory, which is a peculiar theory given its simple groundstate, we have worked mainly on the lattice. 
A possible non-perturbative approach would be to consider the R\'enyi EE of a wider region, say a strip in 2d. This is given by the expectation value of a pair of twist operators, which are defect line-operators defined on the left/right boundaries of the strip. The skeletal limit would result from fusing these line operators, or more precisely, a defect operator product expansion.
It would also be worth investigating other entanglement measures, such as the logarithmic negativity, as well as simpler quantities like fluctuations.
Finally, we expect skeletal entanglement to serve as a \mbox{diagnostic} for spatially anisotropic systems, given its directional nature.

\bigskip

%\pagebreak[5]

\noindent\textit{\textbf{Acknowledgements.}}
We thank A.\,Vigeant for collaboration in the early stages of this work, and H.\,Casini for sharing his code for the thin strip coefficient of the scalar CFT.
We are grateful to B.\,Estienne, I.\,A.\,Kov\'acs, A.\,Paramekanti, G.\,Parez and J.-M.\,Stéphan for stimulating discussions. C.B. is supported by a CRM-Simons Postdoctoral Fellowship at the Université de Montréal. W.W.-K. is funded by a Discovery Grant from NSERC, a Canada Research Chair, and a grant from the Fondation Courtois.

\let\oldaddcontentsline\addcontentsline% Store \addcontentsline
\renewcommand{\addcontentsline}[3]{}% Make \addcontentsline a no-op

\bibliographystyle{utphys} 
\providecommand{\href}[2]{#2}\begingroup\raggedright\endgroup

\let\addcontentsline\oldaddcontentsline% Restore \addcontentsline

%%%%%%%%%% Merge with supplemental materials %%%%%%%%%%
\onecolumngrid
\clearpage
\begin{center}
\textbf{\large Supplemental Material: Entanglement of skeletal regions}\\[.45cm]
  Cl\'ement Berthiere$^{1,2}$ \,and William Witczak-Krempa$^{1,2}$\\[.15cm]
  {\itshape \small ${}^1$Universit\'{e} de Montr\'{e}al, C.\,P.\,6128, Succursale Centre-ville, Montr\'eal, QC, Canada, H3C 3J7\\
  ${}^2$Centre de Recherches Math\'{e}matiques, Universit\'{e} de Montr\'{e}al, Montr\'{e}al, QC, Canada, H3C 3J7\\}
{\small (Dated: \today)}\vspace*{-0.55cm}
\end{center}
%%%%%%%%%% Merge with supplemental materials %%%%%%%%%%
%%%%%%%%%% Prefix a "S" to all equations, figures, tables and reset the counter %%%%%%%%%%
\setcounter{equation}{0}
\setcounter{figure}{0}
\setcounter{table}{0}
\makeatletter
\renewcommand{\theequation}{S\arabic{equation}}
\renewcommand{\thefigure}{S\arabic{figure}}
\renewcommand{\bibnumfmt}[1]{[S#1]}
%\renewcommand{\citenumfont}[1]{S#1}
%%%%%%%%%% Prefix a "S" to all equations, figures, tables and reset the counter %%%%%%%%%%

%\def\l@subsection#1#2{}
\addtocontents{toc}{\vspace{-17pt}}
\tableofcontents % works great.  Includes supplement in toc, as it should.
%\renewcommand{\baselinestretch}{1.0}\normalsize

%\bigskip

% In this section we derive results presented in the main text on the EE of skeletal regions for the free complex boson and free Dirac fermion. 

\vspace{-5pt}

\section{Toric code}\lb{SM-toric}

We explain how the entanglement entropy (EE) results for the toric code were obtained. We work with a square lattice with $k\times k$ sites, and $2k^2$ qubits living on the links. The groundstate under consideration is
\begin{align}
    |\xi_{00}\rangle = |G|^{-1/2} \sum_{g\in G} g|0\rangle \,,
\end{align}
where $G$ is the abelian group generated by all the independent star operators. $|0\rangle$
is the state with all spins down, $\sigma^z_i|0\rangle=-|0\rangle$. 
It contains $|G|=2^{k^2-1}$ elements. There is indeed a global constraint since the product of all stars yields the identity. Hamma \emph{et al.}\ \cite{Hamma:2004vdz} showed that the EE (base 2) of an arbitrary subregion $A$ is given by
\begin{align}
    S(A)=\log_2\left(\frac{|G|}{d_A d_B}\right)\,,
\end{align}
where $d_A=|G_A|$ is the cardinality of the subgroup $G_A$ of $G$, which corresponds to the
elements of $G$ that only act on the subregion $A$: $g\in G_A$ if it can be expressed as $g=g_A\otimes 1_B$. $d_B$ is defined analogously. 
When determining $d_A, d_B$, it is useful to remove one of the stars so that we have $k^2-1$ generators. Since the choice of the star does not affect the groundstate, we can remove a star that acts both on $A$ and $B$, so that it does not belong to $G_A$ or $G_B$.

\begin{figure}[b]
\centering\vspace{-25pt}
\includegraphics[]{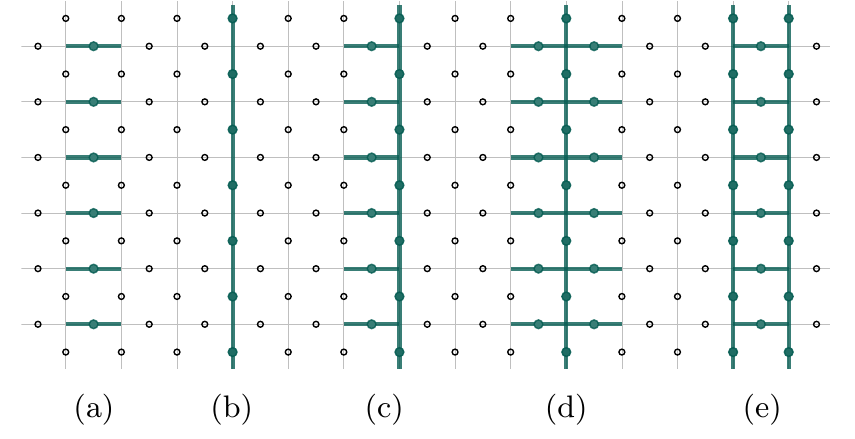}\vspace{-5pt}
\caption{Typical (skeletal) regions on the 2d square lattice for the toric code: (a) Ladder, (b) Chain, (c) Chain-ladder, (d) Ladder-chain-ladder, (e) Chain-ladder-chain.}
\lb{fig4}
\vspace{-5pt}
\end{figure}

\newpage
\noindent\textbf{Ladder.} 
We first consider a ladder that winds around one of the cycles of the torus, as shown in Fig.\,\hyperref[fig4]{\ref{fig4}(a)}.
We remove a star on a vertex touching the ladder.
First, we notice that no operator of $G$ acts purely on $A$, hence $\log_2 d_A=0$. For $G_B$, we have $k(k-2)$ stars that do not touch $A$. In addition, a pair of stars that touch a rung acts only on $B$. We get $k-1$ such pairs, since we need to take into account the removed star. Finally, we recover~\cite{Hamma:2004vdz}
\begin{align}
    S=(k^2-1)-(k^2-2k + k-1) = k\,.
\end{align}

\noindent\textbf{Chain.}
We now consider the vertical bonds that winds entirely around one of the cycles of the torus, as shown in Fig.\,\hyperref[fig4]{\ref{fig4}(b)}.
We remove a star on a vertex touching the chain. Again, we have $\log_2 d_A=0$. For $G_B$, we simply need to remove the $k$ stars that touch $A$: $\log_2 d_B=k(k-1)$. We thus get~\cite{Hamma:2004vdz}:
\begin{align}
    S=(k^2-1)-k(k-1)=k-1\,.
\end{align}

\noindent\textbf{Chain-ladder.}
Let us consider the chain adjacent to the ladder, that winds around one cycle of the torus, as shown in Fig.\,\hyperref[fig4]{\ref{fig4}(c)}. We remove a star that touches the ladder on the left. No star acts only on $A$ so $\log_2 d_A=0$. $G_B$ contains all stars that do not touch $A$: $\log_2 d_B=k^2-2k$. Note that the $2k$ subtraction takes care of the removed star. We thus have
\begin{align}
    S=(k^2-1)-\log_2 d_A -\log_2 d_B = (k^2-1)-(k^2-2k)=2k-1\,.
\end{align}

\noindent\textbf{Ladder-chain-ladder.}
For the ladder-chain-ladder in Fig.\,\hyperref[fig4]{\ref{fig4}(d)}, we remove a star that touches the right side of the ladder. We have $\log_2 d_A=k$ for the middle stars inside $A$. $G_B$ contains $k(k-3)$ stars that do not touch $A$. Note that due to the removed star, the product of all stars that belong to and touch $A$ does not give an element of $G_B$. Thus
\begin{align}
    S=(k^2-1)-k-k(k-3)=2k-1\,.
\end{align}

\noindent\textbf{Chain-ladder-chain.}
Finally, for the chain-ladder-chain as illustrated in Fig.\,\hyperref[fig4]{\ref{fig4}(e)}, we remove a star touching the right chain. Since no element of $G$ acts only $A$, $\log_2 d_A=0$. For $G_B$, we simply need to count the stars that do no touch $A$: $\log_2 d_B=k(k-2)$. No combination of stars that touch $A$ yields an element of $G_B$. We find
\begin{align}
    S=(k^2-1)-(k^2-2k)=2k-1\,.
\end{align}

\noindent\textbf{Topological entanglement entropy from Levin-Wen scheme}.
\begin{figure}[b]
\centering%\vspace{-5pt}
\includegraphics[scale=0.95]{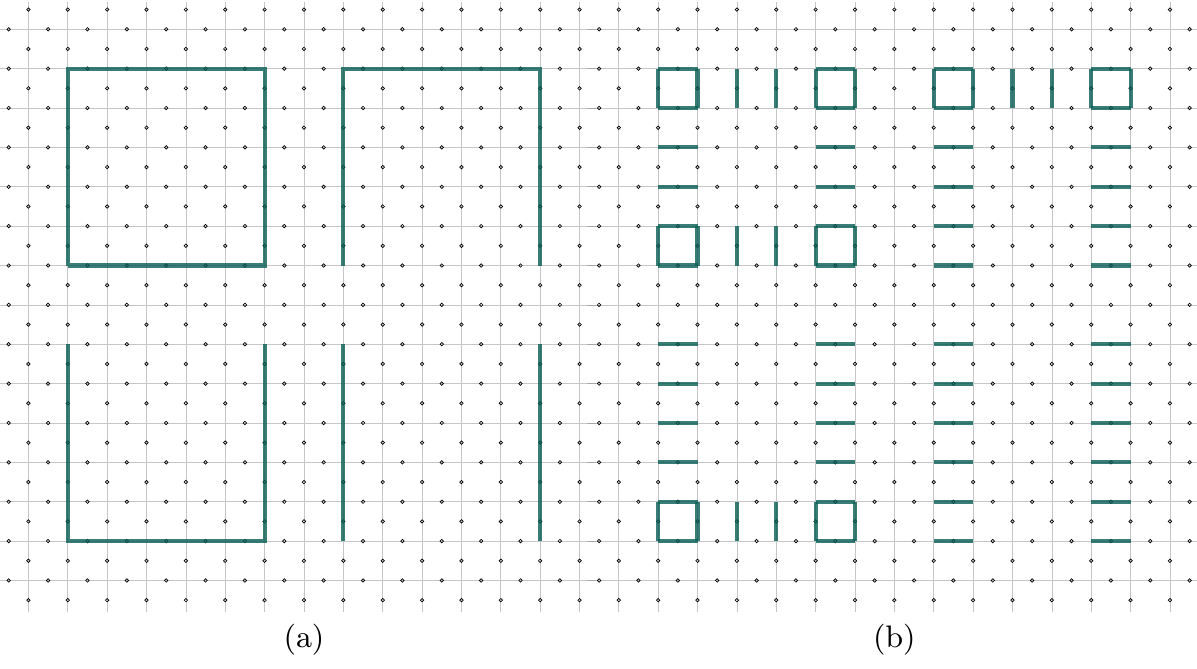}\vspace{-7pt}
\caption{Skeletal regions (in green) on the 2d square lattice for the toric code used in the Levin-Wen \cite{2006PhRvL..96k0405L} subtraction scheme $S(\square)-S(\sqcup) -S(\sqcap)+S(\vcenter{\hbox{\scalebox{1.05}{\hspace{-2.3pt}\rotatebox{90}{$-$}\hspace{-1.8pt}\rotatebox{90}{$-$}}\hspace{-2.5pt}}})$. (a) Chains, and (b) Ladders.}
\lb{fig5}
\end{figure}
We consider skeletal square regions, such as depicted in Fig.\,\ref{fig5}. Using the same method as before we find
\be
S(\square) = n_\square -1 \,,\quad S(\sqcup) = n_\sqcup \,,\quad S(\sqcap) = n_\sqcap \,,\quad S(\vcenter{\hbox{\scalebox{1.05}{\hspace{-2.3pt}\rotatebox{90}{$-$}\hspace{-1.8pt}\rotatebox{90}{$-$}}\hspace{-2.5pt}}}) = n_{\scalebox{0.8}{\hspace{-2pt}\rotatebox{90}{$-$}\hspace{-2pt}\rotatebox{90}{$-$}}}\,,
\ee
for the chains, see Fig.\,\hyperref[fig5]{\ref{fig5}(a)}, such that $S(\square)-S(\sqcup)-S(\sqcap) -S(\vcenter{\hbox{\scalebox{1.05}{\hspace{-2.3pt}\rotatebox{90}{$-$}\hspace{-1.8pt}\rotatebox{90}{$-$}}\hspace{-2.5pt}}})=-1$. Here $n_\square$ is the number of spins in the region, and similarly for the other three regions; we further have $n_\square-n_\sqcup = n_\sqcap - n_{\scalebox{0.8}{\hspace{-2pt}\rotatebox{90}{$-$}\hspace{-2pt}\rotatebox{90}{$-$}}}$. We obtain the same result for the ladders, see Fig.\,\hyperref[fig5]{\ref{fig5}(b)}, and as well as for chain-ladders. Starting from chain-ladder-chains, one has
\be
S(\square) = n_{\partial\square} - 6 \,,\quad S(\sqcup) = n_{\partial\sqcup} -3 \,,\quad S(\sqcap) = n_{\partial\sqcap} -3\,,\quad S(\vcenter{\hbox{\scalebox{1.05}{\hspace{-2.3pt}\rotatebox{90}{$-$}\hspace{-1.8pt}\rotatebox{90}{$-$}}\hspace{-2.5pt}}}) = n_{\partial{\scalebox{0.8}{\hspace{-2pt}\rotatebox{90}{$-$}\hspace{-2pt}\rotatebox{90}{$-$}}}}\hspace{-1.5pt} - 2\,,
\ee
where $n_{\partial\square}$ is the number of spins along the boundary of the region, such that $S(\square)-S(\sqcup) -S(\sqcap)+S(\vcenter{\hbox{\scalebox{1.05}{\hspace{-2.3pt}\rotatebox{90}{$-$}\hspace{-1.8pt}\rotatebox{90}{$-$}}\hspace{-2.5pt}}})=-2$, which is the expected thermodynamic answer \cite{2006PhRvL..96k0405L}. We used the fact 
$n_{\partial\square} -n_{\partial\sqcup} = n_{\partial\sqcap} - n_{\partial\scalebox{0.8}{\hspace{-2pt}\rotatebox{90}{$-$}\hspace{-2pt}\rotatebox{90}{$-$}}}$. 

\newpage
\addtocontents{toc}{\vspace{-5pt}}
\section{Free bosons and Dirac fermions on the lattice}\lb{SM1}

Consider a 2d square lattice, infinite in one direction only, say $x$. We want to compute the EE of a linear chain of $\ell$ sites along the entire transverse direction, say $y$, for the free complex boson and the free Dirac fermion. Given the symmetry, and depending on the boundary conditions imposed in the transverse direction $y$, one can perform a dimensional reduction of the problem.

\subsection{Free bosons}\lb{SM1b}
First, a free complex boson is simply given by two independent free real bosons. We will thus work with real fields, keeping in mind that the EE for the complex boson is given by twice that for the real boson. 
The lattice Hamiltonian of a real massive boson in two spatial dimensions reads
\be\lb{Hb}
H = \frac{1}{2}\sum_{x,y}\Big[ \pi^2_{x,y} + (\phi_{x+1,y}-\phi_{x,y})^2 
 + (\phi_{x,y+1}-\phi_{x,y})^2 + m^2 \phi_{x,y}^2 \Big],
\ee
where the lattice spacing has been set to one, \mbox{$\mathbf{x} = (x,y)$} represents the lattice coordinates with $x\in\mathbb{Z}$ and $y=1, \cdots,\ell$, and $m$ is the mass.
Given the symmetry of the skeletal region, we decompose the fields $\phi_{x,y}$ and $\pi_{x,y}$ along $y$ as
\be
\phi_{x,y} = \sum_{k} f_k(y)\, \phi_x(k)\,,\qquad \pi_{x,y} = \sum_{k} f_k(y)\, \pi_x(k)\,,
\ee
where $f_k(y)$ is a set of orthonormal functions which exact form depend on the boundary conditions imposed along the $y$-direction. %
The Hamiltonian can then be written as a sum over $\ell$ decoupled Hamiltonians
$H = \sum_{k}H_k$,
\be
H_k = \frac{1}{2}\sum_{x}\Big[ \pi^2_{x}(k) + (\phi_{x+1}(k)-\phi_{x}(k))^2 + m_k^2\phi_{x}^2(k) \Big],
\ee
where each $H_k$ corresponds to a 1d free boson with effective mass $m_{k}^2 = m^2+4\sin^2(k/2)$. The quantized momentum $k$ depends on the boundary conditions:
 \be\lb{BCs}
{\rm Twisted:}\quad k=\frac{2\pi(y-\lambda)}{\ell}\,, \qquad\quad {\rm DD:}\quad k=\frac{\pi y}{\ell+1}\,,\qquad\quad {\rm DN:}\quad  k=\frac{\pi(2y-1)}{2\ell+1}\,,\quad
\ee
with $y=1,\cdots,\ell$, and D/N stand for Dirichlet/Neumann boundary conditions. Twisted boundary condition is defined as $\phi_{x,y+\ell}=e^{-2\pi i \lambda}\phi_{x,y}$.
Strictly speaking, the twist parameter $\lambda$ can only take the values $0$ or $1/2$ for the real boson, which correspond to periodic and anti-periodic boundary conditions, respectively. However, for the complex boson, $\lambda$ is not restricted to these two values, i.e. $\lambda\in(0,1)$, and we will thus keep it arbitrary.

The standard method to compute the entanglement and R\'enyi entropies for Gaussian bosonic states of quadratic Hamiltonians can be found in \cite{Casini:2009sr}. In this approach, one calculates the spectrum of the reduced density matrix from the two-point functions $X_{\mathbf{x}\mathbf{x}'}\equiv\la \phi_\mathbf{x}\phi_{\mathbf{x}'}\ra$ and $P_{\mathbf{x}\mathbf{x}'}\equiv\la \pi_\mathbf{x}\pi_{\mathbf{x}'}\ra$ restricted to the subregion $A$ of interest ($X_A$ and $P_A$), such that the Rényi EEs read
\be
&&S_b = \sum_i \bigg[\Big(\nu_i+\frac{1}{2}\Big)\log\Big(\nu_i+\frac{1}{2}\Big)-\Big(\nu_i-\frac{1}{2}\Big)\log\Big(\nu_i-\frac{1}{2}\Big) \bigg],\\
&&S_b^{(n)} = \frac{1}{n-1}\sum_i \log\bigg[\Big(\nu_i+\frac{1}{2}\Big)^n-\Big(\nu_i-\frac{1}{2}\Big)^n\bigg],
\ee
where $\nu_i$ are the eigenvalues of the matrix $C_A=\sqrt{X_AP_A}$. The correlators for the 1d infinite chain are given by
\be
\la \phi_i\phi_j\ra =  \frac{1}{4\pi}\int_{-\pi}^\pi dx \frac{\cos(x(i-j))}{\sqrt{m_k^2+4\sin^2(x/2)}}\,,\qquad\quad  \la \pi_i\pi_j\ra = \frac{1}{4\pi}\int_{-\pi}^\pi dx \sqrt{m_k^2+4\sin^2(x/2)}\cos(x(i-j)) \,.
\ee
In the case at hand, after dimensional reduction, the EE is thus given by $S= \sum_{k}S_b(k)$,
where $S_b(k)$ is the entropy for the $k^{\rm th}$ mode associated to $H_k$.

Since we are considering a skeletal region consisting of a single chain of lattice sites, the reduced entropies $S_b(k)$ are that of a single sites. The correlation matrix $C_A$ is thus one-dimensional with one entry given by
\be\lb{EVb}
\nu(k) = \frac{1}\pi\sqrt{K(-4/m_k^2)E(-4/m_k^2)}\,,
\ee
where $K(x)$ and $E(x)$ are the complete elliptic integrals of the first and second kind, respectively. In all subsequent calculations, we set the mass $m$ of the field to zero.

\subsection{Free Dirac fermions}

We put the free Dirac fermion of mass $m$ on the infinite lattice cylinder. The transverse direction is compactified with twisted boundary condition, similarly as for the complex boson. The dimensional reduction proceeds just as for the free boson, and the 2d Hamiltonian then consists in a sum of decoupled 1d massive free Dirac fermions, $H=\sum_k H_k$,
\be
H_k = \sum_x \left[-\frac{i}{2}\Big(\Psi_x^\dagger\gamma^0\gamma^1(\Psi_{x+1}-\Psi_x) - {\rm h.c.} \Big) + \Psi_x^\dagger \widehat{m}_k\Psi_x \right],\lb{Hf1d}
\ee
where the ``mass operator'' reads $\widehat{m}_k=\gamma^0 m+\gamma^0\gamma^2m_k$, with $m_k=\sin k_y$ and $k_y=2\pi(y-\lambda)/\ell$. The two-dimensional matrices $\gamma^0$ and $\gamma^j$ are Dirac matrices (e.g. $\gamma^0=\sigma_3$ and $\gamma^j=i\sigma_j$, with $\sigma_j$ the Pauli matrices). 

We again use the method discussed in \cite{Casini:2009sr} which relates the eigenvalues of the reduced density matrix to those of the correlation matrix $C=\la \Psi^\dagger\Psi\ra$ restricted to a region $A$ for Gaussian fermionic states of quadratic Hamiltonians. 
The correlator for the 1d infinite chain is given by
\be
\la \Psi_i^\dagger\Psi_j \ra = \frac{1}2\delta_{ij}+ \frac{1}{4\pi}\int_{-\pi}^\pi dx \frac{\widehat m_k+\gamma^0\gamma^1\sin x}{\sqrt{m^2+m_k^2+\sin^2x}}e^{ix(i-j)}\,.
\ee
The expression for the EE in terms of the eigenvalues $\nu_i$ of $C_A$ reads
\be
&&S_f = -\sum_i \big[ \nu_i\log \nu_i+(1-\nu_i)\log(1-\nu_i) \big]\,,\\
&&S_f^{(n)} = \frac{1}{1-n}\sum_i \log\big[ (1-\nu_i)^n +\nu_i^n \big]\,.
\ee
Finally, after dimensional reduction, the EE is thus given by $S= \sum_{k}S_f(k)$,
where $S_f(k)$ is the entropy for the $k^{\rm th}$ mode associated to $H_k$.
Note that due to the fermion doubling on the lattice, one has to divide the lattice results by 4 to get the entropy corresponding to a Dirac field in the continuum limit.

Since $C$ and $1-C$ have the same spectrum, we only retain the single-site correlator eigenvalue
\be\lb{EVf}
\nu(k) = \frac{1}2 + \frac{1}{\pi} K\big(\hspace{-2pt}-(m^2+m_k^2)^{-1}\big),
\ee
such that the reduced entropy is given by $2S_f(\nu)$. In what follow we set the mass $m$ of the Dirac field to zero.

\addtocontents{toc}{\vspace{-5pt}}
\section{Entanglement entropy for free complex bosons and free Dirac fermions}\lb{SM2}
In what follows we will make extensive use of the Euler-Maclaurin formula to obtain asymptotic expansions for the EEs in the scaling limit $\ell\gg 1$. 
The Euler-Maclaurin formula provides an approximation of sums with integrals and vice versa. Let $\ell$ be a positive natural number and $f(x)$ a real valued continuous function of $x\in [1, \ell]$, then
\be\lb{EM}
\sum_{i=1}^\ell f(i) &=& \int_1^\ell \hspace{-1pt} dx\, f(x) + \frac{f(1)+f(\ell)}2 + \sum_{j=1}^p\frac{B_{2j}}{(2j)!}f^{(2j-1)}(x)\Big|_1^\ell  +R_p\,, %+\sum_{j=1}^p\frac{B_{2j}}{(2j)!}f^{(2j-1)}(x)\Big|_1^\ell
\ee
where the $B_i$'s are Bernoulli's numbers (with $B_1=1/2$), and $R_p$ is the remainder term.

\subsection{Complex bosons}

The EE for the skeletal region of interest is given by
\be
S=2\sum_kS_b(k)=2\sum_k \bigg[\Big(\nu(k)+\frac{1}{2}\Big)\log\Big(\nu(k)+\frac{1}{2}\Big)-\Big(\nu(k)-\frac{1}{2}\Big)\log\Big(\nu(k)-\frac{1}{2}\Big) \bigg],
\ee
where the factor two accounts for the two degrees of freedom of the complex boson, $\nu(k)$ is found in \eqref{EVb} and $m_{k} = 2\vert\sin(k/2)\vert$, with momentum $k$ depending on the boundary conditions (see \eqref{BCs}). In the scaling limit $\ell\gg1$, application of the Euler-Maclaurin formula \eqref{EM} yields an area-law $S = \beta_b 2\ell$ at leading order, where
\be
\beta_b = \frac{1}{\pi} \int_0^{\pi} dk\, S_b(k) \simeq 0.1126233777 \,.\quad
\ee
Note that $\beta_b$ satisfies the bounds \eqref{areabounds}, and that it is a monotonic decreasing function of the Rényi index $n$.

The subleading terms depend on the choice of boundary conditions. For twisted boundary conditions we have
\be
S = 2\sum_{y=1}^\ell S_b\Big(\frac{2\pi(y-\lambda)}{\ell}\Big) \simeq 2\int_1^{\ell} \hspace{-2pt}dy\, S_b\Big(\frac{2\pi(y-\lambda)}{\ell}\Big) + S_b\Big(\frac{2\pi(1-\lambda)}{\ell}\Big)+ S_b\Big(\frac{2\pi(\ell-\lambda)}{\ell}\Big) + \cdots\,.
\ee
In the large $\ell$ limit, we find
\be
&&2\int_1^{\ell} \hspace{-2pt}dy\, S_b\Big(\frac{2\pi(y-\lambda)}{\ell}\Big) = \beta_b2\ell -\frac{1}{2}\log\log \ell  -1 +\log\pi + \O\Big(\frac{1}{\log \ell}\Big) \,,\nn\\
&&S_b\Big(\frac{2\pi(1-\lambda)}{\ell}\Big) \simeq S_b\Big(\frac{2\pi(\ell-\lambda)}{\ell}\Big)  =  \frac{1}{2}\log\log \ell +1 -\log\pi  + \O\Big(\frac{1}{\log \ell}\Big)\,.\qquad
\ee
From the Euler-MacLaurin formula, we obtain that the subleading terms are of the form $c_{-i}^b/(\log \ell)^i$, $i=1,2,\cdots$, where the coefficients $c_{-i}^b$ depend on the twist parameter $\lambda$.
The EE of a (twisted) circle on the infinite cylinder is thus given by
\be
S = \beta_b 2\ell -\gamma -s \,, \qquad \gamma=0\,, \qquad s = \frac{c^b_{-1}}{\log \ell} - \frac{c_{-2}^b}{(\log\ell)^2}+\cdots\,,
\ee
where $c_{-i}^b$ can be calculated numerically. As reported in the main text, we have computed $c_{-1}^b$, finding perfect agreement with relation $c_{-1}^b=\log(2\sin\pi\lambda)$.
We note that $c_{-1}^b$ vanishes for the special twist parameter $\lambda=1/6$. In that case, the next higher-order subleading term in $s$ is then universal; we find $c_{-2}^b\vert_{\lambda=1/6}=0.38\pm0.02$. Numerics suggest that $c_{-2}^b\vert_{\lambda=1/6}$ does not depend on R\'enyi's index.

\medskip
For Dirichlet boundary conditions we have
\be
S = 2\sum_{y=1}^\ell S_b\Big(\frac{\pi y}{\ell+1}\Big) \simeq 2\int_1^{\ell} \hspace{-2pt}dy\, S_b\Big(\frac{\pi y}{\ell+1}\Big)+ S_b\Big(\frac{\pi}{\ell+1}\Big)+ S_b\Big(\frac{\pi \ell}{\ell+1}\Big) + \cdots\,,
\ee
where we find for $\ell\gg1$
\be
2\int_1^{\ell} \hspace{-2pt}dy\, S_b\Big(\frac{\pi y}{\ell+1}\Big) &=&\beta_b 2\ell -\log\log \ell + \O(1) \,,\qquad S_b\Big(\frac{\pi}{\ell+1}\Big) = \frac{1}{2}\log\log \ell +  \O(1)\,, \qquad S_b\Big(\frac{\pi \ell}{\ell+1}\Big)  =  \O(1)\,.\qquad
\ee
In the end we obtain
\be
S = \beta_b 2\ell -\gamma -s \,, \qquad \gamma= \frac{1}{2}\log\log \ell + \O(1)\,, \qquad s = \frac{c^b_{-1}}{\log \ell} - \frac{c_{-2}^b}{(\log\ell)^2}+\cdots\,.
\ee
The subleading but divergent term $\gamma$ can be written as $\gamma=2b_D(\pi/2)\log\log \ell + \O(1)$, where $b_D(\pi/2)=1/4$ is the boundary cusp contribution for each orthogonal intersection between the skeletal region and the boundaries of the system. One can further show that the boundary cusp coefficient $b_D(\pi/2)$ is independent of R\'enyi's index.

\medskip
Finally, for Neumann-Dirichlet boundary conditions we have
\be
S = 2\sum_{y=1}^\ell S_b\Big(\frac{\pi (2y-1)}{2\ell+1}\Big) \simeq 2\int_1^{\ell} \hspace{-2pt}dy\, S_b\Big(\frac{\pi (2y-1)}{2\ell+1}\Big)+ S_b\Big(\frac{\pi}{2\ell+1}\Big)+ S_b\Big(\frac{\pi (2\ell-1)}{2\ell+1}\Big) + \cdots\,,
\ee
where we obtain, $\ell\gg1$
\be
2\int_1^{\ell} \hspace{-2pt}dy\, S_b\Big(\frac{\pi y}{\ell+1}\Big) &=&\beta_b 2\ell -\frac{1}2\log\log \ell + \O(1) \,,\quad\; S_b\Big(\frac{\pi}{2\ell+1}\Big) =  \frac{1}{2}\log\log \ell +  \O(1)\,,\quad\; S_b\Big(\frac{\pi(2\ell-1)}{2\ell+1}\Big) = \O(1)\,.\qquad
\ee
The EE hence reads
\be
S = \beta_b 2\ell -\gamma -s \,, \qquad \gamma= \O(1)\,, \qquad s = \frac{c^b_{-1}}{\log \ell} - \frac{c_{-2}^b}{(\log\ell)^2}+\cdots\,.
\ee
For Neumann-Dirichlet boundary conditions, the skeletal region intersects orthogonally the Dirichlet boundary and the Neumann boundary. The subleading term can thus be written as $\gamma= (b_D(\pi/2)+b_N(\pi/2))\log\log\ell + \O(1)$. We thus deduce $b_N(\pi/2) = -b_D(\pi/2) = - 1/4$.
Both quantities are independent of R\'enyi's index.

\subsection{Dirac fermions}

The EE of a twisted circle on the infinite cylinder for the free Dirac fermion is given by 
\be
S=\frac{1}2\sum_kS_f(k)=-\frac{1}2\sum_k \Big[ \nu(k)\log \nu(k)+(1-\nu(k))\log(1-\nu(k)) \Big],
\ee
where we have incorporated a factor 1/4 to account for the fermion doubling problem on the lattice, $\nu(k)$ is found in \eqref{EVf} and $m_{k} = \sin(2\pi(y-\lambda)/\ell)$, $y=1,\cdots,\ell$. In the scaling limit $\ell\gg1$, application of the Euler-Maclaurin formula \eqref{EM} yields an area-law $S = \beta_f 2\ell$ at leading order, where
\be
\beta_f = \frac{1}{2\pi} \int_0^{\pi/2} dk\, S_f(k) \simeq 0.1060777664\,.\quad
\ee
Note that $\beta_f$ satisfies the bounds \eqref{areabounds}, and that it is a monotonic decreasing function of the Rényi index $n$.

Similarly to what has been done for the complex boson, using Euler-Maclaurin formula complemented by a numerical analysis, we find the entropy to be
\be
S = \beta_f 2\ell -\gamma -s \,, \qquad \gamma= 0\,, \qquad s =c_{-1}^f\frac{\log\ell}{\ell^2} -\frac{c_{-2}^f}{\ell^2} + \cdots\,,
\ee
as  reported  in  the  main  text.

\subsection{Bounds on skeletal area-law coefficient}\lb{ApdxBounds}

We explicitly show that the skeletal area-law coefficients for the free complex boson, Dirac fermion and complex Lifshitz ($z=2$) boson satisfy the bounds \eqref{areabounds}, which we recall for convenience:
\be\label{areaboundsB}
\beta_w/w  \leq \beta \leq w \beta_w - (w-1)\beta_{w+1} \leq \beta_w\,.\;\;
\ee
The coefficient $\beta_w$ corresponds to a skeletal region of widths $w$, with the convention $\beta\equiv\beta_{1}$. We gather in Tables\hspace{3pt}\ref{tab1} and \ref{tab3} the values of $\beta_w$ for widths $w=1, 2, 3$ and for R\'enyi indices $n=1, 2, 3$.
The tightest bounds are obtained for $w=2$ in \eqref{areaboundsB}. For the complex boson, we have $\beta_b=0.112623377739$ such that
\be
0.067434763052 \leq \beta_b \leq 0.128121584544 \,,
\ee
while for a Dirac fermion we find $\beta_f=0.106077766441$ such that
\be
0.067216251066 \leq \beta_f \leq 0.128268630477 \,.
\ee
For the Lifshitz boson we obtain $\beta_{\rm Lif}=0.338491005653$ such that
\be
0.203617227388 \leq \beta_{\rm Lif} \leq 0.388178504476 \,.
\ee
The three above inequalities indeed hold, with tighter upper bounds than lower bounds. Note that the upper bound was obtained using the strong subadditivity of EE, which fails in general for R\'enyi index $n\neq1$.

\begin{table}[h]
\caption{\label{tab1}Area-law coefficient $\beta$ for skeletal regions of different widths $w$, for the free complex boson and Dirac fermion.}
\begin{ruledtabular}
\begin{tabular}{ l c c c c c c }
& \multicolumn{3}{c}{Complex boson} & \multicolumn{3}{c}{Dirac fermion}\vspace{3pt}\\
\cline{2-4} \cline{5-7}
 & $n=1$ & $n=2$ & $n=3$ & $n=1$ & $n=2$ & $n=3$ \vspace{3pt}\\
\colrule
$w=1$ \vspace{-1pt}& $0.112623377739$ & $0.054794815045$ & $0.042777703422$ & $0.106077766441$ & $0.079269007259$ & $0.068509183131$\\
$w=2$ \vspace{-1pt}& $0.134869526104$ & $0.061733268142$ & $0.048108112849$ & $0.134432502133$ & $0.079582618865$ & $0.063281181380$ \\
$w=3$ \vspace{-1pt}& $0.141617467663$ & $0.064528392071$ & $0.050222018462$ & $0.140596373790$ & $0.086096921685$ & $0.070784764676$\\
$w=\infty$ \vspace{-1pt}& $0.154903155945$ & $0.071217507275$ & $0.0553290$ & $0.153257055545$ & $0.092367321558$ & $0.0746095$\\
\end{tabular}
\end{ruledtabular}
\end{table}

\addtocontents{toc}{\vspace{-5pt}}
\section{Relation to the entanglement entropy of a thin strip}\lb{SMstrip}

For continuum CFTs, the EE of a strip of length $\ell$ and width $w$ in the thin strip regime $1\ll w\ll\ell$ is given by \cite{Casini:2009sr}
\be\lb{strip}
S_{\rm strip} = \beta_\infty 2\ell - \kappa\frac{\ell}{w} \,,
\ee
where $\beta_\infty$ is the area-law coefficient in the thermodynamic limit, and $\kappa$ is a universal constant that also characterizes the mutual information of two subregions with parallel faces, at very small distance. Values of $\kappa$ for the free complex boson and Dirac fermion, for different Rényi indices $n$, can be found in \cite{Casini:2009sr,Bueno:2015qya}, which we quote in Table\hspace{3pt}\ref{tab2} for convenience. We have obtained refined numerical results for $\kappa$ for $n=1$ for both the boson and fermion CFTs. Expression \eqref{strip} is also valid for the complex Lifshitz $z=2$ boson, for which $\kappa$ is independent of the Rényi index $n$ \cite{Fradkin:2006mb}: $\kappa_{\rm Lif}=\pi/12$.

Using formula \eqref{strip} for a skeletal strip of width $w$, one would obtain a strict area-law
\be\lb{effectiveEE}
S = \beta^{\rm eff}_w 2\ell\,,
\ee
where the effective area-law coefficient is $\beta^{\rm eff}_w=\beta_\infty-\kappa/(2w)$. Now, $\beta^{\rm eff}_w$ is to be compared with the values of $\beta_w$ for skeletal strips in Table\hspace{3pt}\ref{tab1} (and Table\hspace{3pt}\ref{tab3} for the Lifshitz boson). We report values of $\beta^{\rm eff}_w$ and its relative deviation from $\beta_w$ in Tables\hspace{3pt}\ref{tab2} and \ref{tab3}. We notice that $\beta^{\rm eff}_w$ is very close to $\beta_w$, already for $w=1$ and $w=2$.

By considering the skeletal regions as the singular limit of a thermodynamic strip, we observe that $\beta_w\simeq \beta_\infty-\kappa/(2w)$. It is surprising that this relation accurately describes the skeletal regime near the short-distance cutoff. Alternatively, one can use the skeletal coefficient $\beta_w$ to estimate the universal constant $\kappa$.

\begin{table}[h]
\caption{\label{tab2}Effective area-law coefficient $\beta^{\rm eff}_w$ from \eqref{effectiveEE} for the free complex boson and Dirac fermion. Relative deviation between $\beta^{\rm eff}_w$ and $\beta_w$ in parenthesis. Also reported is the universal coefficients $\kappa$ in the EE of a thin strip.}
\begin{ruledtabular}
\begin{tabular}{ l c c c c c c }
& \multicolumn{3}{c}{Complex boson} & \multicolumn{3}{c}{Dirac fermion}\vspace{3pt}\\
\cline{2-4} \cline{5-7}
 & $n=1$ & $n=2$ & $n=3$ & $n=1$ & $n=2$ & $n=3$ \vspace{3pt}\\
\colrule
$\kappa$ \vspace{-1pt}& $0.079301300$ & $0.0455996$ & $0.037339$ & $0.072210$ & $0.0472338$ & $0.040662$\vspace{2pt}\\
$\beta^{\rm eff}_1$ \vspace{-1pt}& $0.11525250$ $(2.33\%)$ & $0.0484177$ $(11.6\%)$ & $0.036659$ $(14.3\%)$ & $0.117152$ $(10.4\%)$ & $0.0687504$ $(13.3\%)$ & $0.054277$ $(20.8\%)$\\
$\beta^{\rm eff}_2$ \vspace{-1pt}& $0.13507782$ $(0.15\%)$ & $0.0598176$ $(3.10\%)$ & $0.045994$ $(4.39\%)$ & $0.135204$ $(0.57\%)$ & $0.0805588$ $(1.23\%)$ & $0.064443$ $(1.84\%)$\\
$\beta^{\rm eff}_3$ \vspace{-1pt}& $0.14168627$ $(0.05\%)$ & $0.0636175$ $(1.41\%)$ & $0.049105$ $(2.22\%)$ & $0.141222$ $(0.45\%)$ & $0.0844950$ $(1.86\%)$ & $0.067832$ $(4.17\%)$
\end{tabular}
\end{ruledtabular}
\end{table}

\begin{table}[h]
\caption{\label{tab3}Area-law coefficient $\beta$ for skeletal regions of different widths $w$, for the free complex Lifshitz boson for $n=1,2,3$. We also report the effective area-law coefficient $\beta^{\rm eff}_w$ from \eqref{effectiveEE} (relative deviation between $\beta^{\rm eff}_w$ and $\beta_w$ in parenthesis).}
\begin{ruledtabular}
\begin{tabular}{ l c c c c }
& \multicolumn{4}{c}{Lifshitz $z=2$ boson} \vspace{3pt}\\
\cline{2-5}
 & $n=1$ & $n=2$ & $n=3$ & $\beta^{\rm eff}_w$ $(n=1)$ \vspace{3pt}\\
\colrule
$w=1$ \vspace{-1pt}& $0.338491005653$ & $0.217792154952$ & $0.183224743303$ & $0.335639879310$ $(0.84\%)$\\
$w=2$ \vspace{-1pt}& $0.407234454776$ & $0.240802841974$ & $0.199499030319$ & $0.401089726260$ $(1.51\%)$\\
$w=3$ \vspace{-1pt}& $0.426290405076$ & $0.251969640317$ & $0.208020922575$ & $0.422906341910$ $(0.79\%)$\\
$w=\infty$ \vspace{-1pt}& $0.466539573210$ & $0.284870029103$ & $0.236890520357$ & $-$\\
\end{tabular}
\end{ruledtabular}
\end{table}

\addtocontents{toc}{\vspace{-5pt}}
\section{Entanglement entropy of singular skeletal regions for the free complex boson}\lb{SM3}

%We report here numerical results for the EE of skeletal regions presenting singularities, such as cusps.

\subsection{Open segment}
We first consider an open straight segment of length $\ell$ on the infinite 2d square lattice. For such geometry, depending on which convention one chooses for what constitutes the perimeter of the skeletal region on the lattice, the constant $\gamma$ can be polluted by the area-law term and is thus ambiguous. However, the subleading term $s$, see \eqref{sub}, contains an unambiguous universal quantity, namely $c_{-1}^b$. We find that for the open segment $c_{-1}^b=0.25\pm0.01$ for $n=1$. Numerical results for higher $n$ suggest that $c_{-1}^b$ is independent of R\'enyi's index.

\subsection{Bulk and boundary cusps}

Let us now consider skeletal regions presenting bulk or boundary cusps (see Fig.\,\hyperref[figZ]{\ref{figZ}(b)} for an example of bulk cusps).
Bulk cusps contribute an $\O(1)$ term in the EE, $\gamma_{\rm bulk}=\sum_i a_{\rm sk}(\theta_i)$, with bulk function $a_{\rm sk}(\theta)$ %in \eqref{cusps} 
which depends on the opening angle $\theta$, and is such that $a_{\rm sk}(\theta)=a_{\rm sk}(2\pi-\theta)$ by symmetry. The coefficient $c_{-1}^b$ in $s$ for free bosons (see \eqref{sub}) is universal for bulk cusps. % but it is not for boundary cusps. 
As for the open segment, the constant terms in the entropy of such skeletal regions may be polluted by the leading area-law term. We thus focus our attention on the unambiguous universal quantity that is $c_{-1}^b$ for bulk cusps.% and $b_{\rm sk}(\theta)$ for boundary cusps.

We computed numerically the EE on the infinite 2d lattice for several polygonal skeletal regions, such as squares, isosceles triangles and octagons. These shapes correspond to bulk cusps of opening angles $\theta=\pi/4,\,\pi/2$ and $3\pi/4$. Numerical results for $c_{-1}^b$ are reported in Table\hspace{3pt}\ref{tab}, and suggest that the contributions of the cusps are additive, i.e. $-c_{-1}^b=\sum_i \tilde{a}_{\rm sk}(\theta_i)$ where $\tilde{a}_{\rm sk}(\theta)$ is a new bulk cusp function. We find that $\tilde{a}_{\rm sk}(\pi/4)\simeq0.59$, $\tilde{a}_{\rm sk}(\pi/2)\simeq0.14$ and $\tilde{a}_{\rm sk}(3\pi/4)\simeq0.028$. We observe a monotonic decrease from $\theta=\pi/4$ to $\pi$, compatible with the intuition that $\tilde{a}_{\rm sk}$ would diverge as $1/\theta$ for small angle and vanish at $\theta=\pi$ since the cusp disappears.
We also observed that $c_{-1}^b$ does not depend on R\'enyi's index.

A boundary cusp produces a double logarithm in the EE, see \eqref{cusps}, with universal coefficient $b_{\rm sk}(\theta)$. The boundary cusp function depends on the angle $\theta$ as well as on the boundary conditions, and by symmetry $b_{\rm sk}(\theta)=b_{\rm sk}(\pi-\theta)$. We computed $b_{\rm sk}(\theta)$ for angles $\theta=\pi/2$ and $\pi/4$, for Dirichlet ($+$) and Neumann ($-$) boundary conditions. We obtain $b_{\rm sk}(\pi/2)=\pm 1/4$ and $b_{\rm sk}(\pi/4)=\pm (0.35\pm0.01)$. Let us stress that the result for $\theta=\pi/2$ is exact and was derived analytically, as shown in Appendix \ref{SM2}. We also note that $b_{\rm sk}(\theta)$ is independent of the Rényi index $n$. More importantly we remark that for both angles considered, $b_{\rm sk}^{(N)}(\theta)=-b_{\rm sk}^{(D)}(\theta)$, where $N/D$ stand for Neumann/Dirichlet boundary conditions. We believe this relation to hold for any angle $\theta\in(0,\pi)$, as explained in the main text.

\begin{table}[h]
\caption{\label{tab}Universal coefficient $c_{-1}^b$ computed numerically for polygonal skeletal regions for the free complex boson.}
\begin{ruledtabular}
\begin{tabular}{ l c c c c }
 & Square & L-shape & Isosceles rectangle triangle & Octagon \vspace{3pt}\\
\colrule
$-c_{-1}^b$ \vspace{-1pt}& $0.57\pm0.01$ & $0.85\pm0.03$ & $1.30\pm0.02$ & $0.23\pm0.02$ \\
\end{tabular}
\end{ruledtabular}
\end{table}

\addtocontents{toc}{\vspace{-5pt}}
\section{Deformations of the lattice free boson}\lb{SM4}

We consider two one-parameter families of deformations of the lattice Hamiltonian \eqref{Hb} for the free boson. Since we want to compute the EE of skeletal regions that allow dimensional reductions, we present the deformed 1d effective Hamiltonians, of the form $H=H_0+H_g$, with
\be
&&H_0 = \frac{1}{2}\sum_{x}\Big[ \pi^2_{x} + (\phi_{x+1}-\phi_{x})^2 + m_k^2\phi_x^2\Big],\\
&&H_g^{(1)} = \frac{g}{2}\sum_{x}(\phi_{x+2}-\phi_{x})^2, \qquad g\ge0\,,\\
&&H_g^{(2)} = \frac{g}{2}\sum_{x}\Big[3(\phi_{x+1}-\phi_{x})^2 
 - (\phi_{x+2}-\phi_{x})^2 \Big],  \qquad 0\le g \le 1\,.
\ee
The first deformation $H_g^{(1)}$ is a simple next-to-nearest neighbor one, while the second $H_g^{(2)}$ interpolates between the relativistic boson ($g=0$) with dynamical exponent $z=1$ and the $z=2$ Lifshitz boson ($g=1$).

The EEs are calculated in the same manner as in Subsection \ref{SM1b}, using the correlators
\be
\la \phi_i\phi_j\ra =  \frac{1}{4\pi}\int_{-\pi}^\pi dq \frac{1}{\omega_{q,k}^{(1,2)}}\cos(q(i-j))\,,\qquad\quad  \la \pi_i\pi_j\ra = \frac{1}{4\pi}\int_{-\pi}^\pi dq \,\omega_{q,k}^{(1,2)}\cos(q(i-j)) \,,
\ee
where
\begin{align}
\omega_{q,k}^{(1)} &= \sqrt{\big(m_k^{(1)}\big)^2 + 4\sin^2(q/2) + 4g\sin^2(q)} \,, &&m_k^{(1)} = 2\sqrt{\sin^2(k/2) + g\sin^2(k)} \,,\\
\omega_{q,k}^{(2)} &= \sqrt{\big(m_k^{(2)}\big)^2 + 4\big(1+g-2g\cos(q)\big)\sin^2(q/2)} \,, \quad &&m_k^{(2)} = 2\sqrt{(1+g-2g\cos(k))\sin^2(k/2)} \,,
\end{align}
with $k$ depending on the boundary conditions along the transverse $y$ direction, see \eqref{BCs}. 
At coincident points $i=j$, the above integrals can be computed in closed forms, though we refrain from showing the results here as they are quite lengthy.

\end{document}